\documentclass[conference]{IEEEtran}
\usepackage{color}
\usepackage{graphicx, pstool}
\usepackage{amsmath}
\usepackage{amsfonts}
\usepackage{amssymb}
\usepackage{booktabs}
\usepackage{dsfont}
\usepackage{cases}
\usepackage{tabularx}
\usepackage{boxedminipage}
\usepackage{url}
\usepackage{mathrsfs}

\begin{document}

\title{\huge On the  Capacity of the Two-Hop Half-Duplex Relay Channel  }
\author{ Nikola Zlatanov$^\dag$, Vahid Jamali$^\ddag$, and Robert Schober$^\ddag$ \\
\IEEEauthorblockA{ $^\dag$ University of British Columbia, Vancouver, Canada, and 
$^\ddag$ Friedrich-Alexander-University Erlangen-N\"{u}rnberg, Germany} 
 \vspace*{-7mm}}

\maketitle

\begin{abstract}
Although extensively investigated, the capacity of the two-hop half-duplex (HD) relay channel is not fully    understood. In particular, a capacity  expression which can be evaluated straightforwardly is not available and an explicit coding scheme which achieves the capacity is not known either. In this paper,   we derive a new expression for the   capacity of the two-hop HD relay channel based on a simplified converse. Compared to previous results,  this capacity expression can be easily evaluated. Moreover, we propose an explicit coding scheme which  achieves the capacity. To achieve the capacity,    the relay does not only send  information   to the destination by   transmitting  information-carrying symbols but also      with the zero symbols resulting from the relay's silence during reception.
As   examples, we compute the capacities of the two-hop HD relay channel  for the cases when  the source-relay and  relay-destination links are  both binary-symmetric channels (BSCs) and additive white Gaussian noise (AWGN) channels, respectively, and   numerically  compare  the capacities  with the rates achieved by conventional relaying where the relay receives and transmits  in a codeword-by-codeword fashion and switches between  reception and transmission in a strictly alternating manner. Our numerical results  show that the capacities of the two-hop HD relay channel  for BSC and AWGN links are significantly larger than the rates achieved with conventional relaying.
\end{abstract}


\IEEEpeerreviewmaketitle

\newtheorem{theorem}{Theorem}
\newtheorem{corollary}{Corollary}
\newtheorem{remark}{Remark}
\newtheorem{example}{Example}
\newtheorem{lemma}{Lemma}
\section{Introduction}
The two-hop relay channel is comprised of a source, a relay, and a destination, where the direct link between  source and destination is not available. In this channel, the message from the source  is first transmitted to the relay, which then forwards it to the destination. Generally, a relay  can employ two different modes of reception and transmission, i.e., the full-duplex (FD) mode and the half-duplex (HD) mode.
 Given the  limitations of current radio implementations, ideal FD relaying is not possible due to  self-interference. The capacity of the two-hop FD relay channel  without self-interference has been derived in  \cite{cover}. On the other hand, although extensively investigated, the capacity of the two-hop HD relay channel is not fully   understood. The reason for this is that   a capacity  expression which can be straightforwardly evaluated is not available and   an explicit coding scheme which achieves the capacity is not known either. Currently, for HD relaying, explicit coding schemes exist only for rates  which are strictly smaller than the capacity,    see \cite{1435648} and \cite{cheap_paper}. To achieve the  rates given in \cite{1435648} and \cite{cheap_paper},   the HD relay   receives a codeword in one time slot, decodes the received codeword, and  re-encodes and re-transmits the decoded information  in the following time slot. However, such fixed switching between reception and  transmission at the relay  was shown to be suboptimal in \cite{kramer2004models}. In particular,  in  \cite{kramer2004models},  it  was shown that if the fixed scheduling of reception and transmission at the HD relay   is abandoned, then additional information can be encoded in the relay's reception and transmission switching pattern yielding an increase in data rate. In addition,   it was shown  in \cite{kramer2004models}  that  the HD relay channel can be analyzed using the framework developed  for the  FD relay channel in \cite{cover}. In particular,   results derived for the FD relay channel   in \cite{cover} can be directly applied to the HD relay channel. Thereby, using the converse  for the degraded relay channel in \cite{cover}, the capacity of the two-hop  HD relay channel is obtained as \cite{kramer2004models}, \cite{kramer_book_1} 
\begin{eqnarray}\label{p-1}
    C=\max_{p(x_1,x_2)}\min\big\{ I(X_1;Y_1|X_2)\;,\; I(X_2;Y_2)\},
\end{eqnarray}
where $X_1$ and $X_2$ are the inputs at source and relay, respectively,  $Y_1$ and $Y_2$ are the outputs at relay and destination, respectively, and $p(x_1,x_2)$ is the joint probability mass function (PMF) of $X_1$ and $X_2$. Moreover, it was shown in \cite{kramer2004models} and \cite{kramer_book_1}  that $X_2$ can be represented as $X_2=[X_2',U]$, where $U$ is an auxiliary random variable with two outcomes $t$ and $r$ corresponding to the HD relay transmitting and receiving, respectively. Thereby, (\ref{p-1}) can be written equivalently as \cite{kramer2004models, 6763001}
\begin{eqnarray}\label{p-2}
    C=\max_{p(x_1,x_2',u)}\min\big\{ I(X_1;Y_1|X_2',U)\;,\; I(X_2',U;Y_2)\},
\end{eqnarray}
where $p(x_1,x_2',u)$ is the joint PMF of $X_1$, $X_2'$, and $U$.
However, the capacity expressions in (\ref{p-1}) and (\ref{p-2}), respectively, cannot be evaluated since it is not known how $X_1$ and $X_2$  nor $X_1$, $X_2'$, and $U$ are mutually dependent, i.e., $p(x_1,x_2)$ and $p(x_1,x_2',u)$ are not known. In fact, the authors of \cite[page 2552]{6763001} state that: ``\textit{Despite knowing the capacity expression (i.e., expression  (\ref{p-2})), its actual evaluation is elusive as it is not clear what the optimal input distribution $p(x_1,x_2',u)$ is.}''
 On the other hand, for the  coding scheme that would achieve (\ref{p-1}) and (\ref{p-2}) if $p(x_1,x_2)$ or $p(x_1,x_2',u)$ was known, it can be argued  that it has to be a decode-and-forward strategy since  the two-hop HD relay channel belongs to the class of  the degraded relay channels defined in \cite{cover}.  Thereby,   the HD relay should  decode  any received codewords, map the decoded information to  new codewords, and transmit  them to the destination. Moreover, it is known from  \cite{kramer2004models}  that such a coding scheme requires the HD
relay   to switch between reception and transmission in a symbol-by-symbol manner, and not in a codeword-by-codeword manner as in \cite{1435648} and \cite{cheap_paper}. However, since    $p(x_1,x_2)$  and $p(x_1,x_2',u)$ are not known and since an explicit  coding scheme does not exist, it is currently not known how to evaluate (\ref{p-1}) and (\ref{p-2}) nor how to encode   additional information  in the relay's reception and transmission switching pattern and thereby  achieve (\ref{p-1}) and (\ref{p-2}).

 Motivated by the above discussion, in this paper, 
we derive a new expression for the   capacity of the two-hop HD relay channel based on a simplified converse. In contrast  to previous results,  this capacity expression can be easily evaluated. Moreover, we propose an explicit coding scheme which     achieves the capacity. 
In particular, we show that achieving the capacity requires the
relay indeed to switch between reception and transmission in a
symbol-by-symbol manner as predicted in \cite{kramer2004models}. Thereby,    the relay does not only send  information   to the destination by   transmitting  information-carrying symbols but also      with the zero symbols resulting from the relay's silence during reception. In addition, we propose a modified coding scheme for practical implementation  where  the HD relay receives and transmits at the same time as in FD relaying, however, the simultaneous reception and transmission is performed such that  self-interference is fully avoided.
  As  examples, we compute the capacities of the two-hop HD relay channel for the cases when the source-relay and  relay-destination links are  both binary-symmetric channels (BSCs) and  additive white Gaussian noise (AWGN) channels, respectively, and we numerically  compare  the capacities with the rates achieved by conventional relaying where the relay receives and transmits  in a codeword-by-codeword fashion and switches between  reception and transmission in a strictly alternating manner. Our numerical results  show that the  capacities of the two-hop HD relay channel for BSC and AWGN links are significantly larger than the rates achieved with conventional relaying.


\section{System Model}\label{sec_2}

The two-hop HD relay channel   consists of a source, a HD relay, and a destination, and the direct link between  source and destination is not available. Due to the HD constraint, the relay cannot transmit and receive at the same time.   In the following, we formally define the channel model.

\subsection{Channel Model} 

The discrete memoryless two-hop HD relay channel is defined by $\mathcal{X}_1$, $\mathcal{X}_2$,     $\mathcal{Y}_1$, $\mathcal{Y}_2$,   and $p(y_1,y_2|x_1,x_2)$, where $\mathcal{X}_1$ and $\mathcal{X}_2$ are the finite  input alphabets at the encoders of the source and the relay, respectively,    $\mathcal{Y}_1$ and $\mathcal{Y}_2$ are the finite  output alphabets at  the decoders of the relay and the destination, respectively,   and $p(y_1,y_2| x_1,x_2)$ is the  PMF  on   $\mathcal{Y}_1\times\mathcal{Y}_2$  for given $x_1\in \mathcal{X}_1$ and $x_2\in \mathcal{X}_2$.  The channel is memoryless in the sense  that given the input symbols   for the $i$-th channel use, the $i$-th output symbols  are independent from all previous input symbols. As a result, the conditional PMF $p(  y_{1}^n, y_{2}^n| x_{1}^n, x_{2}^n)$, where the  notation $a^n$ is   used   to denote the ordered sequence  $a^n=(a_{1}, a_{2},..., a_{n})$, can be factorized as $
     p(  y_{1}^n, y_{2}^n| x_{1}^n, x_{2}^n) =\prod_{i=1}^n p(y_{1i}, y_{2i}| x_{1i}, x_{2i}).$

For the considered channel and the $i$-th channel use, let  $X_{1i}$ and $X_{2i}$ denote the random variables (RVs) which model the input at  source and relay, respectively, and let $Y_{1i}$ and $Y_{2i}$  denote the RVs which model the output at the relay and destination, respectively. 

In the following, we model the HD constraint of the relay and discuss its effect on some important PMFs that will be used throughout this paper.

\subsection{Mathematical Modelling of the HD Constraint}

Due to the HD constraint  of the relay,   the  input  and   output symbols of the relay cannot take non-zero values at the same time. More precisely, for each channel use, if the input symbol of the relay is non-zero then the output symbol has to be zero, and vice versa, if the output symbol of  the relay is non-zero then the input symbol has to be zero. Hence, the following holds
\begin{equation}\label{eq_Y_rv}
     Y_{1i}=\left\{ \hspace{-2mm}
\begin{array}{ll}
Y_{1i}' &  \hspace{-1mm} \textrm{if } X_{2i}=0 \\
0 &  \hspace{-1mm} \textrm{if } X_{2i}\neq 0 .
\end{array} \right.
\end{equation}
where  $Y_{1i}'$ is an RV that take values from the set  $\mathcal{Y}_{1}$.

In order to model the HD constraint of the relay more conveniently, we represent the input set of the relay  $\mathcal{X}_2$ as the union of two sets $\mathcal{X}_2=\mathcal{X}_{2R}\cup \mathcal{X}_{2T}$, where $\mathcal{X}_{2R}$ contains only one element, the zero symbol, and $\mathcal{X}_{2T}$ contains all symbols in $\mathcal{X}_2$ except the zero symbol. Note that, because of the HD constraint, $\mathcal{X}_2$ has to contain the zero symbol.
Furthermore,   we  introduce an auxiliary random variable, denoted by $U_i$, which takes values from the set $\{t,r\}$, where $t$ and $r$ correspond to the relay transmitting a non-zero symbol and a zero symbol, respectively. Hence, $U_i$ is defined as
\begin{equation}\label{eq_U_i}
     U_i=\left\{ \hspace{-2mm}
\begin{array}{ll}
r &  \hspace{-1mm} \textrm{if } X_{2i}=0 \\
t &  \hspace{-1mm} \textrm{if } X_{2i}\neq 0 .
\end{array} \right.
\end{equation}
Let us denote  the probabilities of the relay transmitting a non-zero and a zero symbol for the $i$-th channel use as ${\rm Pr}\{U_i=t\}={\rm Pr}\{X_{2i}\neq 0\}= P_{U_i}$ and ${\rm Pr}\{U_i=r\}={\rm Pr}\{X_{2i}= 0\}=1-P_{U_i}$, respectively.  
We now use (\ref{eq_U_i}) and represent  $X_{2i}$    as a  function  of the outcome of $U_i$. Hence, we have
\begin{eqnarray}\label{eq_X_2i}
     X_{2i}=\hspace{-1mm}\left\{ \hspace{-2mm}
\begin{array}{ll}
0 & \hspace{-2mm} \textrm{if } U_i=r \\
V_{i} & \hspace{-2mm}\textrm{if } U_i=t ,
\end{array}\hspace{-1.5mm} \right. 
\end{eqnarray}
where $V_{i}$ is an RV with distribution $p_{V_i}(x_{2i})$ that takes values from the set  $\mathcal{X}_{2T}$, or   equivalently, an    RV which takes values from the set  $\mathcal{X}_{2}$, but with $p_{V_i}(x_{2i}=0)=0$.
From (\ref{eq_X_2i}), we obtain  
\begin{align}
p(x_{2i}|U_i=r)&=\delta(x_{2i}),\label{eq_rt1}\\
p(x_{2i}|U_i=t)&=p_{V_i}(x_{2i}),\label{eq_rt1a}
\end{align}
where $\delta(x)=1$ if $x=0$ and $\delta(x)=0$ if $x\neq 0$. Furthermore, for the derivation of the capacity,
  we will also need the conditional PMF  $p(x_{1i}|x_{2i}=0)$  which is the  input distribution at the source when   relay transmits a zero (i.e., when $U_i=r$)   and  $p(x_{2i}|U_i=t)= p_{V_i}(x_{2i})$ which is the input distribution at the   relay  when the relay transmits non-zero symbols. As we will see in Theorem~\ref{SItheo_1}, these distributions have to be optimized in order to achieve the capacity. 
 Using $p(x_{2i}|U_i=r)$ and $p(x_{2i}|U_i=t)$, and the law of total probability, the PMF of $X_{2i}$, $p(x_{2i})$, is obtained as
\begin{eqnarray}\label{eq_p(x_2)}
    p(x_{2i}) &=&p(x_{2i}|U_i=t) P_{U_i}+ p(x_{2i}|U_i=r)(1-P_{U_i})\nonumber\\
&\stackrel{(a)}{=} &p_{V_i}(x_{2i}) P_{U_i}+ \delta(x_{2i})(1-P_{U_i}),
\end{eqnarray}
where   $(a)$ follows from (\ref{eq_rt1}) and (\ref{eq_rt1a}). In addition, we will  also need the distribution of $Y_{2i}$, $p(y_{2i})$, which, using the  law of total probability,   can be written  as
\begin{eqnarray}\label{eqe_1}
    p(y_{2i})=  p(y_{2i}|U_i=t) P_{U_i}+ p(y_{2i}|U_i=r) (1-P_{U_i}).
\end{eqnarray}
On the other hand, using $X_{2i}$ and the  law of total probability, $p(y_{2i}|U_i=r)$ can be written as
\begin{align}
&p(y_{2i}|U_i=r) = \sum_{x_{2i}\in \mathcal{X}_2}p(y_{2i},x_{2i}|U_i=r)\nonumber\\
&= \sum_{x_{2i}\in \mathcal{X}_2}p(y_{2i}|x_{2i}, U_i=r) p(x_{2i}|U_i=r)\nonumber\\
 &\stackrel{(a)}{=}   
\sum_{x_{2i}\in \mathcal{X}_2} p(y_{2i}|x_{2i}, U_i=r) \delta(x_{2i}) =  
p(y_{2i}|x_{2i}=0, U_i=r)\nonumber\\ 
 &\stackrel{(b)}{=}  p(y_{2i}|x_{2i}=0)   \label{eq_py2_r},
\end{align}
where $(a)$ is due to (\ref{eq_rt1}) and   $(b)$ is the result of conditioning on the same variable twice since if $X_{2i}=0$ then $U_i=r$, and vice versa.  
On the other hand, using $X_{2i}$ and the law of total probability, $p(y_{2i}|U_i=t)$ can be written as
\begin{align}
   p(y_{2i}|U_i=t)& = \sum_{x_{2i}\in \mathcal{X}_2}p(y_{2i},x_{2i}|U_i=t) \nonumber\\
&= \sum_{x_{2i}\in\mathcal{X}_{2}}  p(y_{2i}|x_{2i}, U_i=t) p(x_{2i}|U_i=t) \nonumber\\
& \stackrel{(a)}{=}  \sum_{x_{2i}\in\mathcal{X}_{2T}}  p(y_{2i}|x_{2i}) p_{V_i}(x_{2i}) 
\label{eq_py2_t},
\end{align}
where $(a)$ follows for (\ref{eq_rt1a}) and since $V_{i}$  takes values from  set $\mathcal{X}_{2T}$. In (\ref{eq_py2_t}),  $p(y_{2i}|x_{2i})$ is the distribution at the output of the  relay-destination channel conditioned on  the relay  transmitting the symbol $x_{2i}$.

\subsection{Mutual Information and Entropy}\label{sec-H(Y_2)}

For the capacity expression given later in Theorem~\ref{SItheo_1}, we need  $I(X_1;Y_1|X_2=0)$, which is the mutual information between the source's input $X_1$ and the relay's output $Y_1$ conditioned on the relay  having its input  set to $X_2=0$, and $I(X_2;Y_2)$, which is the mutual information between the relay's input $X_2$ and the destination's output $Y_2$.

The mutual information $I(X_1;Y_1|X_2=0)$ is obtained by definition as
\begin{align}\label{er_ksndk}
 I\big(X_{1}; Y_{1}|& X_2=0 \big) =\sum_{x_1\in \mathcal{X}_1} \sum_{y_1\in \mathcal{Y}_1} p(y_1|x_1,x_2=0 ) \nonumber\\
&\times p(x_1| x_2=0 ) \log_2\left( \frac{p(y_1|x_1,x_2=0 )}{p(y_1|x_2=0 )} \right),
\end{align}
where   
\begin{align}\label{sdswe}
p(y_1|x_2=0)=\sum_{x_1\in \mathcal{X}_1} p(y_1|x_1,x_2=0 ) p(x_1|x_2=0) .
\end{align}
In (\ref{er_ksndk}) and (\ref{sdswe}), $p(y_1|x_1,x_2=0)$ is the distribution at the output of the source-relay channel conditioned on  the relay  having its input  set to $X_2=0$, and conditioned  on  the input symbols at the source  $X_1$.  

On the other hand, in order to obtain $I(X_2;Y_2)$, we  use the following identity
\begin{equation}\label{eq_ident}
    I(X_2;Y_2)=H(Y_2)-H(Y_2|X_2),
\end{equation}
where    $H(Y_2)$ is the entropy of RV $Y_2$, and $H(Y_2|X_2)$ is the   entropy of $Y_2$ conditioned on the knowledge of $X_2$. The entropy $H(Y_2)$ can be found by definition as 
\begin{eqnarray}\label{eq_H(y2)dad}
    &&\hspace{-8mm}H(Y_2)=-\sum_{y_{2}\in\mathcal{Y}_2}  p(y_{2}) \log_2(p(y_{2})) \nonumber\\
&&\hspace{-7mm}\stackrel{(a)}{=} -\sum_{y_{2}\in\mathcal{Y}_2} \big[ p(y_{2}|U=t) P_{U}+ p(y_{2}|U=r) (1-P_{U}) \big] \nonumber\\
&&\times\log_2\big[ p(y_{2}|U=t) P_{U}+ p(y_{2}|U=r) (1-P_{U})\big], \qquad
\end{eqnarray}
where $(a)$ follows from (\ref{eqe_1}). Now, inserting $p(y_{2}|U=r)$ and $p(y_{2}|U=t)$ given in (\ref{eq_py2_r}) and (\ref{eq_py2_t}), respectively, into (\ref{eq_H(y2)dad}), we obtain the final expression for $H(Y_2)$, as 
\begin{align}\label{eq_H(y2)}
&   H(Y_2) = -\sum_{y_{2}\in\mathcal{Y}_2} \bigg[ P_{U} \sum_{x_{2}\in\mathcal{X}_{2T}}  p(y_{2}|x_{2}) p_{V}(x_{2}) \nonumber\\
&\hspace{28mm}+ p(y_{2}|x_{2}=0) (1-P_{U}) \bigg] \nonumber\\
&\times \log_2\bigg[ P_{U} \hspace{-3mm}\sum_{x_{2}\in\mathcal{X}_{2T}}   \hspace{-3mm} p(y_{2}|x_{2})p_{V}(x_{2}) + p(y_{2}|x_{2}=0) (1-P_{U}) \bigg]. 
\end{align}
On the other hand, the conditional  entropy $H(Y_2|X_2)$ can be found  based on its definition as
\begin{align}\label{eq_H(y2|x2)}
   & H(Y_2|X_2) =-\sum_{x_{2}\in\mathcal{X}_2} p(x_2) \sum_{y_{2}\in\mathcal{Y}_2}  p(y_{2}|x_{2}) \log_2(p(y_{2}|x_2))\nonumber\\
 &\stackrel{(a)}{=}-P_U\sum_{x_{2}\in\mathcal{X}_{2T}} p_{V}(x_{2}) \sum_{y_{2}\in\mathcal{Y}_2}  p(y_{2}|x_{2}) \log_2(p(y_{2}|x_2))\nonumber\\
&\quad \;- (1-P_U) \sum_{y_{2}\in\mathcal{Y}_2}  p(y_{2}|x_{2}=0) \log_2(p(y_{2}|x_2=0)), 
\end{align}
where $(a)$ follows by inserting $p(x_2)$  given in (\ref{eq_p(x_2)}). 
Inserting   $H(Y_2)$ and $H(Y_2|X_2)$ given in (\ref{eq_H(y2)}) and (\ref{eq_H(y2|x2)}), respectively, into (\ref{eq_ident}), we obtain the final expression for $I(X_2;Y_2)$, which is dependent on $p(x_2)$, i.e., on $p_{V}(x_{2})$ and $P_U$. To emphasize the dependance of  $I(X_2;Y_2)$ on $P_U$, we sometimes write $I(X_2;Y_2)$ as $I(X_2;Y_2)\big |_{P_U}$.

We are now ready to present the capacity of the considered channel.

\section{Capacity}

In this section, we provide an easy-to-evaluate expression for the capacity of the two-hop HD relay channel, an explicit coding scheme  that achieves the capacity, and the converse for the capacity.

\subsection{The Capacity}

A new expression for the capacity  of the two-hop HD relay channel is given in the following theorem.

\begin{theorem}\label{SItheo_1}
The capacity of the two-hop HD  relay channel  is given by
\begin{align}\label{SIeq_capacity_2}
    C  =\max_{P_U}\min\Big\{ & \max_{p(x_{1}|x_2=0)} \hspace{-0.5mm} I\big(X_{1}; Y_{1}| X_2=0\big)(1-P_{U}), \nonumber\\
 &  \max_{p_{V}(x_{2})}   I(X_2; Y_{2})\big|_{P_{U} }  \Big\},
\end{align}
where $I\big(X_{1}; Y_{1}| X_2=0 \big)$ is given in (\ref{er_ksndk}) and  $I(X_2; Y_{2})$ is given in (\ref{eq_ident})-(\ref{eq_H(y2|x2)}). The optimal $P_U$ that maximizes the capacity in  (\ref{SIeq_capacity_2}) is given by $P_U^*=\min\{P_U',P_U''\}$, where $P_U'$ and $P_U''$ are the solutions of
\begin{align}\label{eq_PU'}
   \max_{p(x_{1}|x_2=0)} \hspace{-0.5mm} I\big(X_{1}; Y_{1}| X_2=0\big)(1-P_{U})=  \max_{p_{V}(x_{2})}   I(X_2; Y_{2})\big|_{P_{U} }   
\end{align}
and
\vspace{-4mm}
\begin{align}\label{eq_PU''}
  \frac{\partial  \Big(\max\limits_{p_{V}(x_{2})}   I(X_2; Y_{2})\big|_{P_{U} } \Big)}{\partial P_U}=0,
\end{align}
respectively. If $P_U^*=P_U'$, then both terms inside the $\min\{\cdot\}$ function of the capacity in  (\ref{SIeq_capacity_2}) become identical. Whereas, if $P_U^*=P_U''$, then the capacity   in  (\ref{SIeq_capacity_2}) simplifies to
\begin{align}\label{SIeq_capacity_2a}
    C  =    \max_{p_{V}(x_{2})}   I(X_2; Y_{2})\big|_{P_{U}=P_U'' }    = \max_{p(x_{2})}   I(X_2; Y_{2}) ,
\end{align}
which is the capacity of the relay-destination channel.
\end{theorem}
\begin{IEEEproof}
The proof of the capacity  given in (\ref{SIeq_capacity_2}) is provided in two parts. In the first part, given  in Section~\ref{sec_ach}, we show that there exists a coding scheme that achieves a rate $R$ which is smaller, but arbitrarily close to   capacity $C$.  In the second part, given in Section~\ref{sec-converse}, we prove that any rate $R$ for which the probability of error can be made  arbitrarily small, must be smaller than  capacity $C$  given in (\ref{SIeq_capacity_2}). The rest of the theorem follows from solving (\ref{SIeq_capacity_2}) with respect to $P_U$, and simplifying the result. In particular, note that the first  term inside the  $\min\{\cdot\}$ function in (\ref{SIeq_capacity_2}) is a decreasing   function  with respect to $P_U$. This function achieves its maximum for $P_U=0$ and its minimum, which is zero, for $P_U=1$. On the other hand,    the second term inside the  $\min\{\cdot\}$ function in (\ref{SIeq_capacity_2}) is a concave function   with respect to $P_U$. To see this, note that $I(X_2; Y_{2})$ is a concave function with respect to $p(x_2)$, i.e., with respect to the vector comprised of the  probabilities in $p(x_2)$, see \cite{cover2012elements}.  Now, since $1-P_U$ is just the probability $p(x_2=0)$ and  since $p_V(x_2)$ contains the rest of the probability constrained parameters  in $p(x_2)$,  $I(X_2; Y_{2})$ is a jointly concave  function with respect to $p_{V}(x_{2})$ and $P_U$.  In \cite[pp. 87-88]{Boyd_CO}, it is proven that if $f(x,y)$ is a jointly concave function in both $(x,y)$ and $\mathcal{C}$ is a convex nonempty set, then the function $g(x)=\underset{y\in\mathcal{C}}{\max} f(x,y)$ is concave in $x$. Using this result, and noting that the domain of $p_V(x_2)$ is specified by the probability constraints, i.e., by a convex non-empty set, we can directly conclude that $\underset{p_V(x_2)}{\max} \, I(X_2,Y_2)\big|_{P_U}$ is concave with respect to $P_U$.   

Now, the maximization of the minimum  of the decreasing and concave factions with respect to $P_U$, given in (\ref{SIeq_capacity_2}),   has a solution $P_U=P_U''$, when the concave function reaches its maximum, found from (\ref{eq_PU''}), and when for this point, i.e., for  $P_U=P_U''$, the decreasing function is larger than the concave function. Otherwise, the solution is $P_U=P_U'$ which is the point when the decreasing and concave functions intersect,   which is found from (\ref{eq_PU'}) and in which case $P_U'<P_U''$ holds. We note that (\ref{eq_PU'}) has only one solution since for $P_U=1$ the left term in  (\ref{eq_PU'}) becomes zero. Whereas, for $P_U=0$, $\max\limits_{p_{V}(x_{2})}   I(X_2; Y_{2})\big|_{P_{U} =0}= 0$, since  $p(x_2=0)=1$ occurs,  and for $P_U=1$, $\max\limits_{p_{V}(x_{2})}   I(X_2; Y_{2})\big|_{P_{U} =1}\geq 0$, where equality holds if and only if  for $P_U=1$, $p_{V}(x_{2})$ becomes a degenerate (or deterministic) PMF.
\end{IEEEproof}
%

\subsection{Achievability of the Capacity}\label{sec_ach}

In the following, we describe a method for transferring $n R$ bits of information in $n+k$ channel uses,  where $n,k\to\infty$   and   $n/(n+k)\to 1$ as $n,k\to\infty$. As a result, the information is transferred at rate $R$.  To this end, the transmission is carried out in   $N+1$ blocks, where $N\to\infty$. In each block, we use the channel $k$ times. The numbers  $N$ and $k$ are chosen such that $n=Nk$ holds.


We transmit  message $W$, drawn uniformly from  message set $\{1,2,...,2^{nR}\}$, from the source via the HD relay to the destination. To this end, before the start of   transmission,   message $W$ is spilt into $N$ messages, denoted by $w(1),...,w(N)$, where each $w(i)$, $\forall i$, contains $kR$ bits.   The transmission is carried out in the following manner. In   block one,   the source sends message  $w(1)$   in $k$ channel uses to the relay and the relay is silent.  In  block $i$, for $i=2,...,N$, source and relay send  messages $w(i)$ and $w(i-1)$ to relay and destination, respectively,   in $k$ channel uses. In block $N+1$, the relay sends  message  $w(N)$   in $k$ channel uses to the destination and the source is silent. Hence, in the first block and in the $(N+1)$-th block the relay and the source are silent, respectively, since in the first block the relay does not have information to transmit, and in block $N+1$, the source   has no more information to transmit. In blocks $2$ to $N$, both   source and relay transmit, while meeting the HD constraint in every channel use.  Hence, during the $N+1$ blocks, the channel is used $k(N+1)$ times to send $nR=NkR$ bits of information, leading to an   overall information rate   given by
$
  \lim\limits_{N\to\infty} \lim\limits_{k\to\infty}  \frac{ Nk R}{k(N+1)}=R \;\;\textrm{ bits/use}.
$

A detailed description of the proposed coding scheme is given in  the following, where we explain the  rates, codebooks, encoding, and decoding  used for  transmission.

\textit{Rates: } The transmission rate  of  both source and relay is denoted by $R$ and  given by 
\begin{equation}
   R=C    -\epsilon ,\label{SIeq_r_2} 
\end{equation}
 where  $C$ is given in Theorem 1  and $\epsilon>0$ is an arbitrarily small number. 
Note that $R$ is  a function of $P_U^*$, see  Theorem~\ref{SItheo_1}.

\textit{Codebooks: }
We have two codebooks: The source's transmission  codebook and the relay's transmission  codebook.

The source's transmission codebook is generated by mapping each possible binary sequence comprised of $k R$ bits, where $R$ is given by (\ref{SIeq_r_2}),  to a codeword\footnote{The subscript $1|r$ in $\mathbf{x}_{1|r}$  is used to indicate that   codeword $\mathbf{x}_{1|r}$ is comprised of symbols which are  transmitted by the source only when   $U_i=r$.}  $\mathbf{x}_{1|r}$   comprised of $k (1-P_U^*)$  symbols. The symbols in each codeword  $\mathbf{x}_{1|r}$  are generated independently according to   distribution  $p(x_{1}|x_2=0)$. Since in total  there are $2^{k R}$ possible binary sequences comprised of  $k R $  bits, with this mapping we generate $2^{k R }$  codewords $\mathbf{x}_{1|r}$ each  containing $k (1-P_U^*)$ symbols.  These $2^{k R }$  codewords  form the source's transmission codebook, which we  denote    by $\mathcal{C}_{1|r}$.

The relay's transmission  codebook is generated by mapping each possible binary sequence comprised of $k R $ bits, where $R $ is given by (\ref{SIeq_r_2}),  to a  transmission codeword  $\mathbf{x}_2$   comprised of $k$  symbols. The $i$-th symbol, $i=1,...,k$, in  codeword   $\mathbf{x}_2$ is generated in the following manner. For each symbol a coin is tossed. The coin is such that it produces   symbol $r$ with probability $1-P_U^*$ and   symbol $t$ with probability $P_U^*$. If the outcome of the coin flip is $r$, then the  $i$-th symbol of the relay's transmission codeword $\mathbf{x}_2$ is set to zero. Otherwise, if the outcome of the coin flip is $t$, then the $i$-th symbol of  codeword  $\mathbf{x}_2$  is generated independently according to   distribution  $p_{V}(x_{2})$.   The  $2^{k R}$  codewords $\mathbf{x}_2$   form the relay's transmission codebook denoted  by $\mathcal{C}_2$.

The two codebooks are known at all three nodes.

\textit{Encoding,  Transmission, and Decoding: }
In the first block, the source   maps $w(1)$  to the appropriate codeword $\mathbf{x}_{1|r}(1)$ from its codebook $\mathcal{C}_{1|r}$. Then, codeword $\mathbf{x}_{1|r}(1)$ is transmitted  to the relay, which is scheduled to always  receive and be silent (i.e., sets its input to zero)  during the  first block. However,  knowing that the transmitted codeword from the source $\mathbf{x}_{1|r}(1)$ is comprised of $k (1-P_U^*)$ symbols, the relay constructs the received codeword, denoted by $\mathbf{y}_{1|r}(1)$, only from the first  $k (1-P_U^*)$  received symbols.   
In \cite[Appendix~A]{ext_paper}, we prove that  codeword $\mathbf{x}_{1|r}(1)$ sent in the first block can be decoded successfully from the received codeword at the relay $\mathbf{y}_{1|r}(1)$ using a typical decoder \cite{cover2012elements} since $R $ satisfies
\begin{eqnarray}\label{eq_d_1aa}
   R  < \max_{p(x_{1}|x_2=0)} I\big(X_{1}; Y_{1}| X_2=0\big) (1-P_U^*).
\end{eqnarray}

In blocks $i=2,...,N$, the encoding,  transmission, and decoding are performed as follows. In    blocks $i=2,...,N$, the source and the relay map $w(i)$ and $w(i-1)$ to the appropriate codewords $\mathbf{x}_{1|r}(i)$ and $\mathbf{x}_2(i)$ from codebooks $\mathcal{C}_{1|r}$ and $\mathcal{C}_{2}$, respectively. Note that the source also knows  $\mathbf{x}_2(i)$ since  $\mathbf{x}_2(i)$ was generated from $w(i-1)$ which the source transmitted in the previous (i.e., $(i-1)$-th)   block.
The transmission of  $\mathbf{x}_{1|r}(i)$ and $\mathbf{x}_2(i)$ can be performed in two ways: 1) by the relay  switching between reception and transmission, and 2) by the relay always receiving and transmitting as in FD relaying. We first explain the first option.

Note that both  source and   relay know the position of the zero symbols in $\mathbf{x}_2(i)$. Hence, if the  first symbol  in  codeword $\mathbf{x}_2(i)$ is zero, then in the first symbol interval of block $i$, the source transmits its first symbol from codeword  $\mathbf{x}_{1|r}(i)$ and the relay receives. By receiving, the relay actually also sends the first symbol of codeword $\mathbf{x}_2(i)$, which is the symbol zero, i.e., $x_{21}=0$.   On the other hand, if the first symbol  in   codeword $\mathbf{x}_2(i)$  is non-zero, then in the first symbol interval of block $i$, the relay transmits its first symbol from codeword  $\mathbf{x}_2(i)$ and the source is silent. 
 The same procedure  is performed   for the $j$-th channel use in block $i$, for $j=1,...,k$. In particular,  if the  $j$-th symbol  in  codeword $\mathbf{x}_2(i)$ is zero, then in the $j$-th channel use of block $i$  the source transmits its next untransmitted symbol from codeword  $\mathbf{x}_{1|r}(i)$ and the relay receives. With this reception, the relay actually also sends the $j$-th symbol of codeword $\mathbf{x}_2(i)$, which is the symbol zero, i.e., $x_{2j}=0$. On the other hand, if the $j$-th symbol  in   codeword $\mathbf{x}_2(i)$ is non-zero, then for the $j$-th channel use of block $i$, the relay transmits the $j$-th symbol of codeword  $\mathbf{x}_2(i)$ and the source is silent. 
Since   codeword $\mathbf{x}_2(i)$ contains approximately $k(1-P_U^*)$ symbols zeros,    for $k\to\infty$ the source   can transmit practically all\footnote{Due to the strong law of large numbers, the number of zero symbols in  $\mathbf{x}_2(i)$ is  $k(1-P_U)- \varepsilon$, where $\varepsilon$ is an integer satisfying $\lim\limits_{k\to\infty} \varepsilon/k=0$ \cite{cover2012elements}. Hence, when we say  that the source can transmit practically all of its symbols, we mean either all or all except for a negligible fraction  $\lim\limits_{k\to\infty} \varepsilon/k=0$ of them. This fraction is negligible such that the decisions  of the typical decoder  are not affected for $k\to\infty$, see \cite[Appendix~A]{ext_paper}.} of its $k(1-P_U^*)$ symbols from codeword $\mathbf{x}_{1|r}(i)$ during a single block to the relay. Let  $\mathbf{y}_{1|r}(i)$ denote the corresponding received   codeword at the relay.   In \cite[Appendix~A]{ext_paper}, we prove that the codewords $\mathbf{x}_{1|r}(i)$ sent in blocks $i=2,\dots,N$ can be decoded successfully at the relay from the corresponding received codewords $\mathbf{y}_{1|r}(i)$  using a typical decoder \cite{cover2012elements} since   $R$ satisfies (\ref{eq_d_1aa}). 
On the other hand, the relay  sends  the entire codeword $\mathbf{x}_2(i)$, comprised of  $k$ symbols of which approximately $k(1-P_U^*)$ are zeros,  to the destination. In particular, the relay sends the  $k(1-P_U^*)$ zero symbols of codeword $\mathbf{x}_2(i)$ to the destination by being silent during reception, and sends the remaining $kP_U^*$  symbols of codeword $\mathbf{x}_2(i)$ to the destination by actually transmitting them.   On  the other hand, the destination   listens during the entire block and   receives a codeword  $\mathbf{y}_2(i)$.   In \cite[Appendix~B]{ext_paper}, we prove that the destination  can successfully decode  $\mathbf{x}_2(i)$ from the received codeword $\mathbf{y}_2(i)$, and thereby obtain $w(i-1)$, since   rate $R$ satisfies  
\begin{eqnarray}
     R <\max_{p_{V}(x_{2})} I(X_2; Y_{2})\Big|_{P_{U}=P_{U}^* }  .     \label{SIeq_r_2a}
\end{eqnarray}

In a practical implementation,  the relay may not be able to switch between reception and transmission in a symbol-by-symbol manner, due to  practical constraints   regarding the speed of switching. Instead, we may allow the relay to receive and transmit at the same time and in the same frequency band similar to FD relaying. However, this simultaneous reception and transmission is performed while  avoiding self-interference  since, in each symbol interval,  either the  input or the  output information-carrying symbol   of the relay is zero. 
This is accomplished   in the following manner. The source performs the same operations  as for the case   when the relay switches between reception and transmission. On the other hand, the relay transmits all symbols from $\mathbf{x}_2(i)$ while continuously listening. Then, the relay discards from the received codeword, denoted by  $\mathbf{y}_1(i)$,   those symbols for which the corresponding symbols in $\mathbf{x}_2(i)$ are non-zero, and only collects the symbols in $\mathbf{y}_1(i)$ for which the corresponding symbols in    $\mathbf{x}_2(i)$ are equal to zero. From the collected symbols in $\mathbf{y}_1(i)$, the relay obtains the received information-carrying  codeword $\mathbf{y}_{1|r}(i)$ which it needs for decoding. Codeword $\mathbf{y}_{1|r}(i)$  is completely free of self-interference since the symbols in $\mathbf{y}_{1|r}(i)$  were received in symbol intervals for which the corresponding transmit symbol at the relay was zero

In the last (i.e., the $(N+1)$-th) block, the source is silent and the relay transmits $w(N)$ by mapping it to the corresponding   codeword   $\mathbf{x}_2(i)$ from set $\mathcal{C}_{2}$. The relay transmits all   symbols in codeword   $\mathbf{x}_2(i)$ to the destination. The destination  can decode the received codeword in block $  N+1 $ successfully, since (\ref{SIeq_r_2a}) holds.

Finally, since both  relay and destination can decode their respective codewords  in each block, the entire message $W$ can be decoded successfully at the destination at the end of the $(N+1)$-th block.

\subsection{Converse}\label{sec-converse}
As shown in \cite{kramer2004models}, the HD relay channel can be investigated with the framework  developed for the FD relay channel in \cite{cover}. Since the considered two-hop HD relay channel belongs to the class of degraded relay channels  defined in \cite{cover}, the rate of this channel is upper bounded    by  \cite{cover}, \cite{kramer2004models}
\begin{eqnarray}\label{eq_1_0a1}
 &&\hspace{-7mm}    R\leq \max_{p(x_1,x_2)}  \min \big\{  I\big(X_{1}; Y_{1}|X_{2}\big) ,    I\big(X_{2};Y_{2}\big)\big\}  .
\end{eqnarray}
On the other hand, $I\big(X_{1}; Y_{1}|X_{2}\big)$ can be simplified as
\begin{align}\label{rav_1aa}
 & I\big(X_{1}; Y_{1}|X_2\big)\nonumber\\
  &  =   I\big(X_{1}; Y_{1}| X_{2}=0\big) (1-P_{U}) +   I\big(X_{1}; Y_{1}|X_{2}\neq 0 \big) P_{U} \nonumber\\
&  
 \stackrel{(a)}{=}
   I\big(X_{1}; Y_{1}| X_{2}=0\big) (1-P_{U}),  
\end{align}
where    $(a)$ follows from (\ref{eq_Y_rv}) since  when $X_{2}\neq 0$,  $Y_{1}$ is deterministically zero thereby leading to $I\big(X_{1}; Y_{1}|X_{2}\neq 0\big)= 0$. 
 Inserting  (\ref{rav_1aa}) into (\ref{eq_1_0a1}),  (\ref{eq_1_0a1})  simplifies as
\begin{equation}\label{eq_1_0a2}
    R\leq \max_{p(x_1,x_2)}   \min \big\{  I\big(X_{1}; Y_{1}| X_{2}=0\big) (1-P_{U})   \; ,\;    I\big(X_{2};Y_{2}\big)\big\}  .
\end{equation}
Since $p(x_1,x_2)=p(x_1|x_2)p(x_2)$, where $p(x_2)$ is given in (\ref{eq_p(x_2)}) as a function of $P_U$ and $p_V(x_2)$, the maximization in (\ref{eq_1_0a2}) with respect to $p(x_1,x_2)$ can be resolved into joint maximization with respect to $p(x_1|x_2)$, $p_V(x_2)$, and $P_U$. Now, since $I\big(X_{1}; Y_{1}| X_{2}=0\big) (1-P_{U})  $ and  $I\big(X_{2};Y_{2}\big)$  are functions of $p(x_1|x_2=0)$ and $p_V(x_2)$, respectively, and no other function  inside the $\min\{\cdot\}$ function  in (\ref{eq_1_0a2}) is dependent on the distributions  $p(x_1|x_2=0)$ and $p_V(x_2)$,   (\ref{eq_1_0a2})   can be written equivalently  as
\begin{align}\label{eq_1_0a}
R\leq  \max_{P_U} \min \Big\{ & \max_{p(x_1|x_2=0)} I\big(X_{1}; Y_{1}| X_{2}=0\big) (1-P_{U})  \; ,\nonumber\\
&  \max_{p_V(x_2)}  I\big(X_{2};Y_{2}\big)\Big\},
\end{align}
where    $\max\limits_{p(x_1|x_2=0)} I\big(X_{1}; Y_{1}| X_{2}=0\big)$ and  $\max\limits_{p_V(x_2)}  I\big(X_{2};Y_{2}\big)$ exist   since these functions are concave  with respect to $p(x_1|x_2=0)$ and $p_V(x_2)$, respectively.
On the other hand, the maximum in (\ref{eq_1_0a}) with respect to $P_{U}$ exists since the first and the second terms inside the $\min\{\cdot\}$ function in (\ref{eq_1_0a}) are   monotonically decreasing and   concave functions with respect to $P_U$, respectively (see proof of Theorem~\ref{SItheo_1} for concavity). This concludes the proof of the converse.

\section{Application Examples: BSC and AWGN}\label{sec_num}
In this section, we use Theorem~1 to derive the capacity of the two-hop HD relay channel for the cases when the   source-relay and relay-destination links are both  BSCs and AWGN channels, respectively. 

\subsection{BSC}
Assume that the source-relay and relay-destination links are both  BSCs with probability of error $P_{\varepsilon 1}$ and $P_{\varepsilon 2}$, respectively.  Let $H(P)=-P\log_2(P)-(1-P)\log_2(1-P)$ denote the binary entropy function.
Then, the capacity for this channel is given in the following corollary.
\begin{corollary}\label{cor_1}
The capacity of the considered  relay channel with   BSCs links is given by
 \begin{align}\label{eq_dnfj}
C=&\max_{P_U} \min\{   (1-H(P_{\varepsilon1}))(1-P_U),\nonumber\\
&-A\log_2(A)-(1-A)\log_2(1-A) - H(P_{\varepsilon 2}) \},
\end{align}
where  $
   A=P_{\varepsilon 2}(1-2 P_U)+P_U$,
and is achieved with
 \begin{align} 
 &  p_{V}(x_{2})=\delta(x_2-1) \label{eq_ww2}\\
&  p(x_1=0|x_2=0)= p(x_1=1|x_2=0)  = 1/2\label{eq_ww3}.
\end{align}
There are two cases for the   optimal $P_U$  which maximizes (\ref{eq_dnfj}).
 If $P_U$ found  from
\begin{align} \label{eq_dnfj1}
&(1-H(P_{\varepsilon1}))(1-P_U) \nonumber\\
&= -A\log_2(A)-(1-A)\log_2(1-A) - H(P_{\varepsilon 2})
\end{align}
is smaller than $1/2$, then the optimal $P_U$ which maximizes (\ref{eq_dnfj}) is found as the solution to (\ref{eq_dnfj1}). In this case, the first and second term inside the $\min\{\cdot\}$ of the capacity become equal.
Otherwise, if $P_U$ found from (\ref{eq_dnfj1}) is $P_U\geq 1/2$, then the optimal $P_U$ which maximizes (\ref{eq_dnfj}) is $P_U=1/2$, and the capacity simplifies to
\begin{eqnarray}\label{eq_cap_BSC_1}
    C= 1-H(P_{\varepsilon 2}),
\end{eqnarray} 
which is the capacity of the relay-destination link.
\end{corollary}
\begin{IEEEproof}
Please refer to \cite[Section~IV-A]{ext_paper}.
\end{IEEEproof}


\subsection{AWGN}
We now assume that the source-relay and relay-destination links are AWGN channels, i.e., channels which are impaired by  independent, real-valued, zero-mean AWGN with variances $\sigma_1^2$ and $\sigma_2^2$, respectively. More precisely, the outputs at the relay and the destination are given by
$
    Y_k=X_k+N_k,\quad k\in\{1,2\},
$
where $N_k$ is a zero-mean Gaussian  RV with variance $\sigma_k^2$, $k\in\{1,2\}$, with distribution $p_{N_k}(z)$,  $k\in\{1,2\}$, $-\infty\leq z\leq\infty$.
Moreover, assume that the  symbols transmitted by the source  and the relay must satisfy the following average power constraints\footnote{If the optimal distributions $p(x_1|x_2=0)$ and $p_{V}(x_{2})$ turn out to be continuous,  the sums in (\ref{eq_cond_awgn_power}) should be replaced by integrals.} 
\begin{equation}\label{eq_cond_awgn_power}
   \sum_{x_1\in\mathcal{X}_1} x_1^2\; p(x_1|x_2=0)\leq P_1  \;\textrm{ and } \;\sum_{x_2\in\mathcal{X}_{2T}} x_2^2\; p_{V}(x_{2}) \leq P_2.
\end{equation}
Then, the capacity for this channel is given in the following corollary.
\begin{corollary}\label{cor_2}
The capacity of the considered relay channel where the source-relay and relay-destination links are both  AWGN channels with noise variances $\sigma_1^2$ and $\sigma_2^2$, respectively, and where the average power constraints of the inputs of source and relay are given by (\ref{eq_cond_awgn_power}), is given by
\begin{align}\label{eq_cap_gauss}
    C &= \frac{1}{2} \log_2\left(1+\frac{P_1}{\sigma_1^2}\right)(1-P_U^*)\nonumber\\
&\stackrel{(a)}{=} - \int\limits_{-\infty}^\infty \left( P_U^*\; \sum_{k=1}^K p_k^* p_{N_2}(y_2-x_{2k}^*) + (1-P_U^*) p_{N_2}(y_2)  \right) \nonumber\\
&\hspace{-2mm}\times \log_2\left( P_U^*\; \sum_{k=1}^K p_k^* p_{N_2}(y_2-x_{2k}^*) + (1-P_U^*) p_{N_2}(y_2)  \right)   d y_2 \nonumber\\
&\quad-\frac{1}{2}\log_2(2\pi e \sigma_2^2),  
\end{align}
where  
the optimal $P_U^*$ is found such that  equality $(a)$ in (\ref{eq_cap_gauss}) holds. The capacity in (\ref{eq_cap_gauss}) is achieved when $p(x_1|x_2=0)$ is the zero-mean Gaussian distribution with variance $P_1$ and $p_{V}^*(x_{2})=\sum_{k=1}^K p_k^* \delta(x_2-x_{2k}^*)$ is a discrete distribution  which satisfies
\begin{eqnarray}\label{eq_cond_2}
    \sum_{k=1}^K p_k^* =1\quad \textrm{and}\quad \sum_{k=1}^K p_k^* (x_{2k}^*)^2=P_2.
\end{eqnarray}
 and maximizes $H(Y_2)$. 
\end{corollary}
\begin{IEEEproof}
Please refer to \cite[Section~IV-B]{ext_paper}.
\end{IEEEproof}

Unfortunately, obtaining the optimal $p_{V}(x_{2})$ which satisfies (\ref{eq_cond_2}) and maximizes $H(Y_2)$ in closed form is difficult, if not impossible, see \cite[Section~IV-B]{ext_paper} for more details. Therefore,   a  brute-force search  has to be   used in order to find $x_{2k}^*$ and $p_k^*$, $\forall k$. Instead of an optimal discrete input distribution at the relay $p_{V}^*(x_{2})$,    we can use\footnote{Note that when $p_{V}(x_{2})$ is a continuous Gaussian distribution, the probability that $x_2=0$ will occur  is zero. Hence, the definition of $p_{V}(x_{2})$ is not violated.}  a continuous, zero-mean Gaussian distribution with variance $P_2$, which will produce a rate smaller than the capacity, given by
\begin{align} \label{eq_gauss_num}
 &   R_{\rm Gauss}  = \frac{1}{2} \log_2\left(1+\frac{P_1}{\sigma_1^2}\right)(1-P_U)\nonumber\\
&\stackrel{(a)}{=} - \int\limits_{-\infty}^\infty \big( P_U\; p_{G}(y_2) + (1-P_U) p_{N_2}(y_2)  \big) \nonumber\\
& \hspace{-1mm}\times \log_2 \big( P_U\; p_{G}(y_2)\hspace{-0.5mm} + \hspace{-0.5mm}(1-P_U) p_{N_2}(y_2)  \big)   d y_2\hspace{-0.5mm}  -\hspace{-0.5mm} \frac{1}{2}\log_2(2\pi e \sigma_2^2), 
\end{align}
 where $P_U$ is found such that equality $(a)$ holds and $p_{G}(y_2)$ is a continuous, zero-mean Gaussian distribution with variance $P_2+\sigma_2^2$. 
In the following, we numerically evaluate the capacities in  Corollaries~\ref{cor_1} and ~\ref{cor_2}.

\section{Numerical Examples}
In this section, we  numerically evaluate the capacities of the two-hop HD relay channel when the source-relay and relay-destination links are both BSCs  and AWGN channels, respectively, and compare it to the maximal achievable rates of conventional relaying \cite{1435648}, \cite{cheap_paper}.
 
\subsection{BSC}
\begin{figure} 
\centering
\includegraphics[width=3.75in]{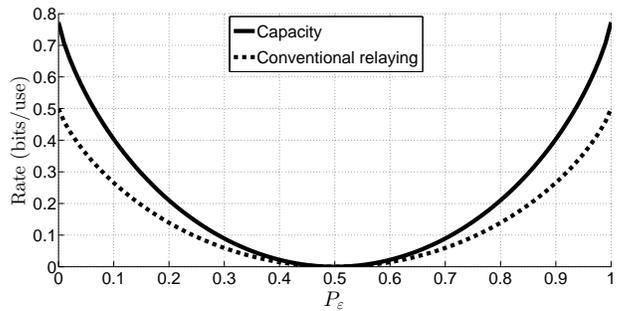}
\vspace*{-3mm}
\caption{Comparison of rates for the BSC as a function of the error probability  $P_{\varepsilon 1}=P_{\varepsilon 2}=P_{\varepsilon}$.}\label{fig_n1}
\vspace*{-1mm}
\end{figure}

For simplicity, we assume symmetric links with $P_{\varepsilon 1}=P_{\varepsilon 2}=P_{\varepsilon}$. The capacity is given by (\ref{eq_dnfj}). This capacity is plotted in Fig.~\ref{fig_n1}, where the optimal $P_U$ is found from (\ref{eq_dnfj1}) using a  mathematical software package, e.g. Mathematica. As a benchmark, in Fig.~\ref{fig_n1}, we also show  the maximal achievable rate using conventional relaying, obtained  as $
    R_{\rm conv}= \big(1- H(P_\varepsilon)\big)/2.$
 As can be seen from Fig.~\ref{fig_n1}, the capacity   is significantly higher than the  maximal rate of conventional relaying. For example, 
 when both links are error-free, i.e., $P_{\varepsilon}=0$,  conventional relaying achieves $0.5$ bits/channel use, whereas the capacity is  $0.77291$, which is   $54\%$ larger than the rate achieved with conventional relaying.  We note that this value was also reported in \cite[page 327]{kramer_book_1}, where only the case of  error-free BSCs was investigated.

\subsection{AWGN}
 
For the AWGN case,  the capacity is evaluated based on Corollary~\ref{cor_2}. However, since for this case the optimal input distribution at the relay $p_V(x_2)$ is unknown, i.e., the values of $p_k^*$ and $x_{2k}^*$ in (\ref{eq_cap_gauss}) are unknown, we have performed a brute force search for the   values of $p_k^*$ and $x_{2k}^*$ which maximize (\ref{eq_cap_gauss}).  Two examples of such distributions  are shown in  \cite[Fig.~6]{ext_paper}  for two different values of the SNR $ P_1/\sigma_1^2= P_2/\sigma_2^2$.  

\begin{figure}
\includegraphics[width=3.75in]{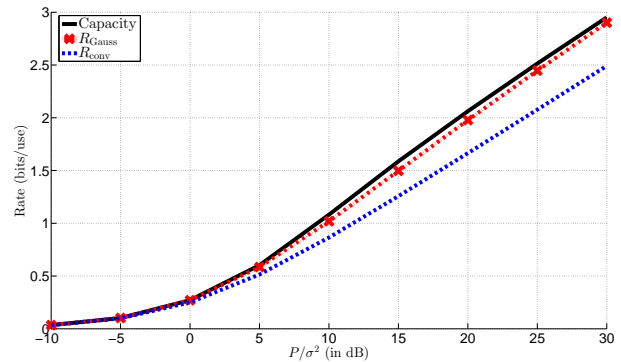}
\centering
\vspace*{-3mm}
\caption{Source-relay and relay destination links are AWGN channels with  $ P_1/\sigma_1^2= P_2/\sigma_2^2=P/\sigma^2$.}\label{fig_n2}
\vspace*{-1mm}
\end{figure}

The capacity is shown in  Fig.~\ref{fig_n2}   for the case when   $P_1/\sigma_1^2=P_2/\sigma_2^2=P/\sigma^2$. In Fig.~\ref{fig_n2}, we also show the rate achieved when instead of an optimal discrete input distribution at the relay $p_V^*(x_2)$,   we use  a continuous, zero-mean Gaussian distribution with variance $P_2$, in which case the rate is given in (\ref{eq_gauss_num}).  From  Fig.~\ref{fig_n2}, we can see that $R_{\rm Gauss}$ is smaller than the capacity, which was expected. However, the loss in performance caused by the Gaussian inputs is moderate, which suggests that the performance gains obtained by the proposed coding scheme are mainly due to the exploitation  of the silent (zero) symbols for conveying information from the HD relay to the destination rather than the optimization of $p_V(x_2)$. 
As benchmark, in Fig.~\ref{fig_n2}, we have also shown the maximal achievable rate using conventional relaying, obtained  for  $P_1/\sigma_1^2=P_2/\sigma_2^2=P/\sigma^2$  as \cite{1435648}, \cite{cheap_paper} $
    R_{\rm conv}= \log_2\left(1+ P/\sigma^2\right)/4 $.
Comparing the capacity   with $R_{\rm conv}$ in Fig.~\ref{fig_n2}, we see that for $10 \textrm{ dB}\leq P/\sigma^2\leq 30$ dB,   $3$ to $5$ dB gain is achieved. Hence, large performance gains are achieved using the proposed capacity coding scheme even if  suboptimal input distributions at the relay are employed.

\section{Conclusions}
We have derived an easy-to-evaluate  expression for the   capacity of the two-hop HD relay channel based on a simplified  converse.   Moreover, we have proposed an explicit coding scheme which   achieves the capacity. In particular, we showed that the capacity is achieved  when additional information is sent by the relay to the destination using the zero symbol implicitly sent by the relay's silence  during reception. Furthermore, we have evaluated the  capacity for the cases  when both links are BSCs and AWGN channels, respectively.  From the numerical examples, we have observed that the capacity    is significantly larger than the rate  achieved with conventional relaying protocols.

\bibliography{litdab}

\begin{thebibliography}{1}
\providecommand{\url}[1]{#1}
\csname url@samestyle\endcsname
\providecommand{\newblock}{\relax}
\providecommand{\bibinfo}[2]{#2}
\providecommand{\BIBentrySTDinterwordspacing}{\spaceskip=0pt\relax}
\providecommand{\BIBentryALTinterwordstretchfactor}{4}
\providecommand{\BIBentryALTinterwordspacing}{\spaceskip=\fontdimen2\font plus
\BIBentryALTinterwordstretchfactor\fontdimen3\font minus
  \fontdimen4\font\relax}
\providecommand{\BIBforeignlanguage}[2]{{%
\expandafter\ifx\csname l@#1\endcsname\relax
\typeout{** WARNING: IEEEtran.bst: No hyphenation pattern has been}%
\typeout{** loaded for the language `#1'. Using the pattern for}%
\typeout{** the default language instead.}%
\else
\language=\csname l@#1\endcsname
\fi
#2}}
\providecommand{\BIBdecl}{\relax}
\BIBdecl

\bibitem{cover}
T.~Cover and A.~{El Gamal}, ``{Capacity Theorems for the Relay Channel},''
  \emph{IEEE Trans. Inform. Theory}, vol.~25, pp. 572--584, Sep. 1979.

\bibitem{1435648}
A.~Host-Madsen and J.~Zhang, ``{Capacity Bounds and Power Allocation for
  Wireless Relay Channels},'' \emph{IEEE Trans. Inform. Theory}, vol.~51, pp.
  2020 --2040, Jun. 2005.

\bibitem{cheap_paper}
M.~Khojastepour, A.~Sabharwal, and B.~Aazhang, ``On the {C}apacity of '{C}heap'
  {R}elay {N}etworks,'' in \emph{Proc. Conf. on Inform. Sciences and Systems},
  Princeton, NJ, 2003.

\bibitem{kramer2004models}
G.~Kramer, ``{Models and Theory for Relay Channels with Receive Constraints},''
  in \emph{Proc. 42nd Annual Allerton Conf. on Commun., Control, and
  Computing}, Monticello, IL, 2004, pp. 1312--1321.

\bibitem{kramer_book_1}
G.~Kramer, I.~Mari{\'c}, and R.~D. Yates, \emph{{Cooperative
  Communications}}.\hskip 1em plus 0.5em minus 0.4em\relax Now Publishers Inc.,
  2006, vol.~1, no.~3.

\bibitem{6763001}
M.~Cardone, D.~Tuninetti, R.~Knopp, and U.~Salim, ``{On the Gaussian
  Half-Duplex Relay Channel},'' \emph{IEEE Trans. Inform. Theory}, vol.~60, pp.
  2542--2562, May 2014.

\bibitem{cover2012elements}
T.~M. Cover and J.~A. Thomas, \emph{{Elements of Information Theory}}.\hskip
  1em plus 0.5em minus 0.4em\relax John Wiley \& Sons, 2012.

\bibitem{Boyd_CO}
S.~Boyd and L.~Vandenberghe, \emph{Convex Optimization}.\hskip 1em plus 0.5em
  minus 0.4em\relax Cambridge University Press, 2004.

\bibitem{ext_paper}
\BIBentryALTinterwordspacing
N.~Zlatanov, V.~Jamali, and R.~Schober, ``{Capacity of the Two-Hop Half-Duplex
  Relay Channel},'' \emph{Extended Version, Submitted to IEEE Trans. Inform.
  Theory}. [Online]. Available: \url{http://arxiv.org/abs/1411.5299}
\BIBentrySTDinterwordspacing

\end{thebibliography}
\bibliographystyle{IEEEtran}
\end{document}